# Mega-ampère to mega-gauss: the generation of intense magnetic fields using fast pulsed-power drivers

P.-A. Gourdain[*,1,2], G. Brent[2], J.B. Greenly[3], D. A. Hammer[3], R. V. Shapovalov[1,2]

[1] Extreme State Physics Laboratory, Physics and Astronomy Department, University of Rochester, NY 14627

[2] Laboratory for Laser Energetics, University of Rochester, NY 14627

[3] Laboratory for Plasma Studies, Cornell University, Ithaca, NY 14850

**Abstract**:

Intense magnetic fields modify the properties of extremely dense matter via complex processes that call for precise measurements in very harsh conditions. This endeavor becomes even more challenging because the generation of mega-gauss fields in a laboratory is far from trivial. This paper presents a unique and compact approach to generate fields above 2 mega-gauss in less than 150 ns, inside a volume close to half a cubic centimeter. Magnetic insulation, keeping plasma ablation close to the wire surface, and mechanical inertia, limiting coil motion throughout the current discharge, enable the generation of intense magnetic fields where the shape of the conductor controls the field topology with exquisite precision and versatility, limiting the need for mapping exactly magnetic fields.

---

[*] Corresponding author: gourdain@pas.rochester.edu

Processes taking place in strongly-coupled, degenerate matter[1] under giga-gauss fields are not well understood as the Coulomb force acts as a small perturbation over the magnetic force[2]. The properties of dense matter are still impacted when the field is on the order of one mega-gauss[3]. These interactions are keys to understanding the physics of neutron star atmospheres or the material properties of planetary cores at the heart of gas giants. In recent years, high-power lasers, pulsed-power generators and heavy-ion beams have reached mass densities and temperatures relevant to explore the exotic properties of matter under extreme conditions. Yet, the external magnetic fields required to overcome the intrinsic fields produced high power drivers must be larger than one mega-gauss[4,5]. Facilities capable of producing magnetic fields strong enough usually lack the drivers required to compress matter in such extreme regimes, and vice versa.

Fortunately, it is not necessary to undertake such research under steady-state magnetic fields as no known material can resists such magnetic pressures. So, large fields have been produced for decades, using pulsed power supplies[6,7]. However, even a transient approach is still problematic. Coils are under immense stresses, and bulky mechanical supports are required to minimize coil deformation. Under such conditions, any design will put severe constraints on the shape of the conductor and limit access to the experimental volume where fields are most concentrated. Indeed, when the structure is too invasive, the pulse length must be reduced, dropping from the millisecond[8] down to the microsecond[9,10,11]. As the pulse length becomes smaller, it is material inertia that keeps the coil together. A single turn coil can produce tens of mega-gauss when driven by a high-power lasers, but only for a couple of nanoseconds[12,13]. The vacuum magnetic energy cannot exceed the laser energy, on the order of a kilo-joule, limiting the experimental volume to several cubic-millimeters. The laser also produces a background plasma, that change the field topology and rob some current from the coil.

This paper presents a novel approach to the generation of mega-gauss fields, using a 100-ns rise-time pulsed-power driver. These drivers can efficiently concentrate a large amount of energy in a volume large enough to accommodate high energy density experiments. Mega-ampere currents provide the magnetic insulation[14,15] that suppresses electrical breakdown, allowing multi-turn coils to be used. The rise time is fast enough to use mass inertia against unwanted motion of the conductor, even with wires only a millimeter across. Without a mechanical structure, the number of field topologies is virtually infinite.

Mega-gauss fields were generated using a modest pulsed-power generator, called the COrnell Beam Research Accelerator[16] (COBRA). The next generation of pulsed-power devices, like HADES[17], offers the same capability in a small footprint (< 2 m), that can be moved to a high-power laser facility or stand next to a heavy ion beam.

The idea to use a pulsed-power driver as a mega-gauss field generator germinated from recent experiments where the return current post was used to generate an axial field to magnetize a plasmas[18,19,20]. Simulations[21,22] have also extended the magnetization capability to a variety to fusion platforms, such as MagLIF[23]. The main contribution this paper brings forth is a clear demonstration that mega-ampere currents can produce mega-gauss magnetic fields with a variety of field topologies, in an environment virtually plasma-free. A volume dominated by plasmas would be quickly filled by magnetic fields of unknown strength and direction. Two cases are explored sequentially to show that a fast-pulsed power diver can be turned into a precise mega-gauss field generator. First, we show how plasma processes affect probes measurements and how a deviation from a controlled field distribution can be detected very effectively. In a second time, we show that a plasma free environment generates fields which topology is solely controlled by the shape of the conductor. Three-dimensional numerical simulations confirmed that the field measured are mostly axial and above 2MG. In conclusion, the computation of the field topology using Biot-Savart law is an excellent surrogate to a three-dimensional experimental field mapping.

## 1 Experimental setup

The experimental setup is shown in Figure 1-a. We used a two-turn solenoid formed by a 1.25mm wire with a 3-mm pitch and a bore diameter of the coil is 6 mm, all measured from edge to edge of the wire. The total experimental volume is just shy of 0.5 cm³. The coil geometry, shown in Figure 1-b, highlights two important advantages brought forth by fast pulses. This geometry cannot be used on longer time scales, as mechanical supports would block access to the experimental volume. This geometry also tests the performance of magnetic insulation. The plasma formed on the wire surface would cross the gap between two consecutive turns a change in field topology would be immediately detected by two miniature Bdot probes[24,25], measuring the axial (Ba) and edge (Be) magnetic fields (also shown in Figure 1-b). Both probes have intrinsic measurements errors, due to alignment and calibration error.

The Bdot probes used in this work are single-turn loops fabricated at the end of a 15-cm length of 0.5 mm OD copper semi-rigid coaxial cable. The cable is totally encapsulated in Kapton tubing sealed at its end with epoxy. This insulation prevents electrical connection of the coax or the loop to ambient plasma. Insulation failure in COBRA is immediately detected; it results in prompt excursion of the signal to kV negative voltages. The small size of the probes allows them to be located where needed, leaves enough separation from high-voltage electrodes to eliminate any significant capacitive pickup, and minimally perturbs any ambient plasma. The loop areas are 0.1- 0.2 mm². These small areas are necessary both for localization of the measurement point, and to keep the output signals below breakdown in the coaxial cable. Signals as high as 800V, and Bdot values approaching 100 T/ns have been recorded without failure. In our experiments, sensitivity was measured to be 10% or less of the axial field sensitivity. The probes are calibrated by insertion into a magnetic field coil driven by a 200ns risetime pulser that gives signals of order 1V. The signal calibration uncertainty is +/-5% but imprecision in orientation of the loop as fabricated limits the experimental uncertainty of measurement of a specified field component to +/- 10%. These probes have been tested in opposite-polarity pairs, and the signals track within the calibration uncertainty.

We are now interested in detectable field variations, caused by plasmas or coil deformation, rather than absolute field strengths and direction. Without plasma and deformation, the field strength and direction can be directly computed using a purely electromagnetic code. Side-on visible laser backlighting, the green beam in Figure 1-a, was used to monitor the progression of the plasma across the gap between coil turns. When the density reaches ~$10^{20}$ cm³, the laser is refracted away from the axis of the optical system and light disappears from the picture at this location. End-on XUV imagining, the purple beam in Figure 1-a, highlights hot plasma regions (~10-100 eV) which may not be dense enough to be detected with the laser backlighter.

## 2 The generation of mega-gauss fields

Magnetized high energy density experiments call for a precise measurement of the field strength and direction to understand how the field impact material properties. Usually experimental samples have small dimensions compared to the skin depth of the field with 100 ns time rise and field deflection caused by eddy currents can be neglected at peak current. The biggest source of field error comes from the plasma. A plasma free environment eliminates completely the need for a three-dimensional mapping of the field. In this spirit, the research presented here dealt with both situations. In a first time, we used an experimental setup where a plasma was produced purposefully to evaluate how it perturbs probe measurements and how it appears on different diagnostics (XUV, shadowgraphy). This experiment allowed to measure the field error caused by the plasma, on top of probe calibration errors. In a second time, we used a setup virtually free of plasma, and show that that field errors are greatly reduced.

### 2.1 The detection of plasma effects on probe measurements

The most direct method to detect the impact of the plasma is measuring the field at two separate locations. The first location is inside the bore of the solenoid, filled by the plasma formed on the wire surface. The second probe is located 10 mm away from the edge of the coil, in a region where the plasma cannot easily interfere with field measurements. Ideally, no probe should be inside the solenoid as it takes valuable experimental space. However, our goal here is only to measure the magnetic field, and having a probe on axis makes sense. In the absence of plasma, the ratio between both fields

$$r_1 = \frac{Ba}{Be} \qquad (1)$$

should remain constant throughout the experiment, providing that there is no coil motion throughout the shot. If a plasma is present near the axial probe, then the ratio will change in time. Initially, for tens of nanoseconds after the current starts to flow, there is not plasma and the baseline ratio can be determined precisely, simply by setting the error

$$e_1 = \frac{Ba - r_1 Be}{Ba_{max}} \qquad (2)$$

to zero. Figure 2-a shows that a ratio $r_1$ of 7.4 sets the error $e_1$ to virtually zero for the first 30 ns of the current discharge. As the experiment progresses, plasma effects (e.g.

electrical breakdown, probe shielding) have caused a build-up of the error $e_1$ up to 7%, at peak current (t~130 ns).

A similar error detection can be undertaken using only one probe. We can compare the field measured by the probe against the discharge current, measured by a Rogowski coil, located 20 cm below the coil and shown in Figure 1-a. We define two other errors:

$$e_2 = \frac{Ba}{r_1 Be_{max}} - \frac{I}{I_{max}} \text{ and } e_3 = \frac{Be}{Be_{max}} - \frac{I}{I_{max}}. \tag{3}$$

Again, in the absence of dynamical effects, both errors should be close to zero. $e_2$ compares the normalized Ba ($\underline{Ba}$) to the normalized current I ($\underline{I}$). The renormalization of Ba uses $r_1 Be_{max}$ rather than $Ba_{max}$, because $Ba_{max}$, as measured, will vary from shot to shot due to dynamical effects. Rather, $r_1 Be_{max}$ is this maximum axial field produced by the coil, as inferred by a measurement devoid of plasma interference. $e_3$ compares the normalized Be ($\underline{Be}$) to $\underline{I}$. Once the parameters $r_1$, $Be_{max}$ and $I_{max}$ are determined, they can be carried from shot to shot, as long as the coil has the same shape and the probe is always placed at the same location. Once the different parameters are defined, namely $r_1 = 7.4$, $Be_{max} = 173$ T and $I_{max} = 796$ kA, they are used in the rest of this paper.

Any dynamical effects will cause $e_2$ and $e_3$ to vary in time. For instance, the plasma ablated from the coil can enter the anode-cathode feed, shown below the coil Figure 1-b, or reach the axial probe. As Figure 2-b shows, the field at both locations tracks the discharge current for the first 30 ns. This is expected, according to Figure 2-a. As time passes, both errors start to deviate away from 0. On the one hand, $e_2$ diverges rapidly, reaching up to 11%. This large error comes from the plasma formed at the coil surface, clearly visible in the extreme UV spectrum shown on Figure 3-a, which has been pushed towards the axis and surrounds the probe (seen using a thresholding routine on Figure 3-b). On the other hand, $e_3$ stays below 2% throughout the discharge, an indication that plasma effects on the edge probe and inside the anode-cathode gap are minor. Figure 3-c plots both axial and edge fields, where the errors are caused by plasma or conductor motion. The axial Bdot probe shows that the field on axis as reached 192 T, with an error of +/-11%. The scaled edge Bdot probe signal gives a field of 173T, with a 2% error. The error $e_3$ is barely visible on the plot.

Overall, the proposed method can successfully detect the impact of dynamical effects on field measurements. While the small error $e_3$ indicates that Be might be an acceptable scaled measurement of the axial field, once the ratio $r_1$ has been determined, the next section shows that the axial probe can be used to measure the field directly in an environment without plasma.

## 2.2 The generation of mega-gauss magnetic fields in plasma-free environments

We now use an experimental setup which does not promote plasma formation together with a single probe on axis, measuring the axial field at the coil center. As Figure 4-a shows, the plasma is now hardly visible on the XUV detectors. While the previous shot had an axial field signal with a large error, the absence of plasma allows to measure the field on axis directly. There is now good agreement between the time evolution of the field and the time evolution of the current. Figure 4-c shows that the field on axis reached 172T for a peak discharge current of 810 kA. The error $e_2$ at this time is 2.1%. The errors at later times come from probe failure, usually detectable on the non-integrated Bdot signal[26]. The laser shadowgraphs of Figure 5, from a different shot, shows the probe will eventually fails since the formation of a plasma is ultimately expected. The backlighter triggered when the gap between two turns of the solenoid just closed, 250 ns after the discharge started. The green laser light cut-off indicates that refraction would preclude a visible high-power laser to reach a target placed at the center of the coil. This is not a problem since a high-power laser would be fired at peak field, 130ns into the discharge, where the gap is still wide opened. Then a higher current shot was performed on COBRA and the current reached a total 1.04 MA, yielding a magnetic field of 224 T, and an error $e_2$ of 1.2%. Figure 4-b shows that more plasma has formed on the conductor surface compared to the low current shot of Figure 4-a.

## 3 Numerical simulations

We now turn to numerical simulations to evaluate the impact of intrinsic errors caused by Bdot probe calibration and positioning. We first start with a semi-analytical code that computes the field, splitting the coil geometry into a series of straight segments of finite length. We can use Biot-and-Savart formula to compute the field from one segment,

$$\vec{B} = \frac{\mu_0}{4\pi a^2}(\cos\theta_1 - \cos\theta_2)\vec{I} \times \vec{a},$$

(4)

and then sum all the fields together to get the total field. Here the vector $\vec{I}$ has the strength and direction of the current flowing inside the wire. $\theta_1$ and $\theta_2$ are the angles between the wire direction and the direction defined by the end of the wires and the location where the field is computed. $\vec{a}$ is the vector defined by the shortest distance between the wire direction and this location. This computation suppose that the currents are steady state inside an infinitely small conductor. Since we compute the field only outside the conductor used in the experiment, the vector $\vec{a}$ is never null, and the field can be computed. Figure 6-a shows the magnetic field strength for the coil geometry used experimentally and computed semi-analytically. The field on axis reaches 260 T for a current of 1.04 MA. This field is 10% higher than the one measured experimentally. Figure 6-b shows the magnetic field lines are mostly axial even if the solenoid does not have turns close together.

However, the analytical code has two basic shortcomings. It assumes that all the current flows along the innermost region of the wire. In reality, the current is distributed throughout the wire cross-section. This distribution is also not homogeneous. It also ignores how the current transitions from the electrodes to the wire. To further evaluate these effects, we used PERSEUS[27], a fully three-dimensional two-fluid magneto-hydrodynamic code, that can capture the dynamics of the system as the current rises. We model large volume, including the anode cathode region of COBRA, to obtain the right boundary condition across the anode-cathode gap. The domain was decomposed in 252 million cells, yielding a resolution of 0.1 mm. The lower limit placed on plasma density was $10^{21} m^{-3}$. Figure 7-a shows the strength of the magnetic field, with iso-contours, at peak current. The cut-off density for the iso-volume following the coil and electrodes shape is $10^{25} m^{-3}$. Above this density laser propagation becomes problematic. As the figure shows, the space between coil turns is large enough to let high-power lasers reach the target inside the solenoid bore. Figure 7-a also shows that numerical simulations are in good agreement with the semi-analytical code. However, all the iso-contours in the MHD simulation have shifted radially inwards. This suggests that some of the current has not gone into the wire, shrinking the iso-countours closer to the coil center. According to the code, the current has fully penetrated the wire at peak current. Non-linear current diffusion, caused by large resistivity gradients[22], tends to overtake linear current diffusion (i.e. skin effect) at such current densities. Figure 7-b shows large temperature variations, responsible for the resistivity gradients behind the non-linear diffusion of the current. The field lines in Figure 7-b shows that the magnetic field is mostly oriented along the solenoid axis, in good agreement with semi-analytical results. Figure 7-c shows that most of the plasma is confined near the wire surface as one would expect when magnetic insulation performs accordingly. The field on axis is 238T, at the coil center. This value is lower than the field computed by the semi-analytical code, yet 6% higher than the field measured experimentally. This error is consistent with the probe and current calibration errors.

## 4 Conclusions

We have shown for the first time that a fast pulsed-power generator can produce mega-gauss fields using wound solenoids. The fast rise time, on the order of 100 ns, reduces the risk of coil deformation during the discharge and magnetic insulation limits the flow of currents outside the conductors. Overall, the field strength, measured using miniature Bdot probes, is in very good agreement with numerical simulations. This mode of operation has several advantages compared to systems with slower rise times. The design of the coil and its construction are simple and inexpensive. This is a big advantage since the coil is destroyed and must be rebuilt after every shot. Short timescales allow to remove all mechanical structures and large currents produce magnetic insulation. With a physical insulator gone, high energy density drivers and diagnostics have direct side-on and end-on lines of sight into the experimental volume. Finally, the magnetic field generated is well inferred with a simple electromagnetic code, where the coil, anode and cathode geometries are explicitly modelled, alleviating the need for a three-dimensional mapping of the field inside the actual solenoid.

By measuring the field at the center of the coil, both experiments and MHD simulations have shown that the plasmas produced are mostly confined to the wire surface. While this work shows that well-designed Bdot probes can reliably measure mega-gauss fields for hundreds of nano-seconds, they carry intrinsic measurement errors, due to the difficulty in evaluating accurately their active area (especially in the presence of pulsed fields, where flux penetration becomes an important player), and how they are positioned with respect to the solenoid. While the experimental data collected with Bdot probes cannot vouch for a one-to-one mapping between numerical codes and experiment, the agreement between field evolution, as measured by the probes and the discharge current shows that parasitic plasma formation in the coil vicinity is negligible. As a result, if the shape of the conductor is the same in every experiment, then the field at a given location will be identical as long as the peak current is the same. Reasonable rise time variations have little impact. Ultimately many different techniques may have to be used to measured fields more accurately via local methods, such as Zeeman splitting[28] or electron cyclotron emission[29]. However, the field topology will only be controlled by the shape of the conductor, giving immense freedom in how dense matter can be magnetized.

## 5 Methods

### 5.1 COBRA

COBRA is a pulsed power generator capable of generating 1 MA with a variable current rise time (from 100 to 200 ns). The generator is powered by two Marx banks, housing 16 1.35 µF capacitors, that were charged up to 70 kV. The total stored energy is 100 kJ and the overall efficiency of the pulse-forming lines is greater than 10%. The driver's low impedance (~0.5 Ω) is achieved by using four identical pulse-forming lines, 1.8 Ω each. The COBRA load current monitor calibration is accurate to +/5%.

### 5.2 Bdot Probes

The Bdot probes used in this work are single-turn loops fabricated at the end of a 15-cm length of 0.5 mm OD copper semi-rigid coaxial cable. The cable is totally encapsulated in Kapton tubing sealed at its end with epoxy. This insulation prevents electrical connection of the coax or the loop to ambient plasma. Insulation failure in COBRA is immediately detected; it results in prompt excursion of the signal to kV negative voltages.

The small size of the probes allows them to be located where needed, leaves enough separation from high-voltage electrodes to eliminate any significant capacitive pickup, and minimally perturbs any ambient plasma. The loop areas are 0.1- 0.2 mm$^2$. These small areas are necessary both for localization of the measurement point, and to keep the output signals below breakdown in the coaxial cable. Signals as high as 800V, and Bdot values approaching 100 T/ns have been recorded without failure. In our experiments, sensitivity was measured to be 10% or less of the axial field sensitivity. The probes are calibrated by insertion into a magnetic field coil driven by a 200ns risetime pulser that gives signals of order 1V. The signal calibration uncertainty is +/-5% but imprecision in orientation of the loop as fabricated limits the experimental uncertainty of measurement of a specified field component to +/- 10%. These probes have been tested in opposite-polarity pairs, and the signals track within the calibration uncertainty.

## 5.3 PERSEUS

The PERSEUS code solves two-fluid MHD equations together with the Generalized Ohm's Law. The additional physics currently in the model goes beyond standard MHD by including the electron inertia, electron pressure and Hall terms. This model is more accurate than the standard MHD model. The electron inertia and Hall physics allows for a consistent treatment of the low density (or vacuum) region and does not require a nonphysical vacuum resistivity. This resistivity de facto controls the plasma ablation at the material interface with vacuum in the code. When electron physics is computed properly the plasma ablation is controlled by energy deposition rather than vacuum resistivity. The code is second order in space using a flux-limited implicit-explicit MUSCL scheme (Monotone Upwind Scheme for Conservation Laws). Due to the implicit time stepping of electron physics, the code runs as fast as standard MHD codes. The numerical simulations were done on the IBM Blue Gene/Q, operated by the University of Rochester.

**Acknowledgements**: The research was supported in part by the NSF under the grant PHY-1725178, the Department of Energy National Nuclear Security Administration under Awards Numbers DE-SC0016252 and DE-NA0001944 and the University of Rochester.

**Authors contributions:**
PAG, JBG, GB, RVS and DAH developed the load for COBRA. PAG, JBG, DAH analyzed the data. PAG developed the Biot-Savart code and ran PERSEUS simulations.

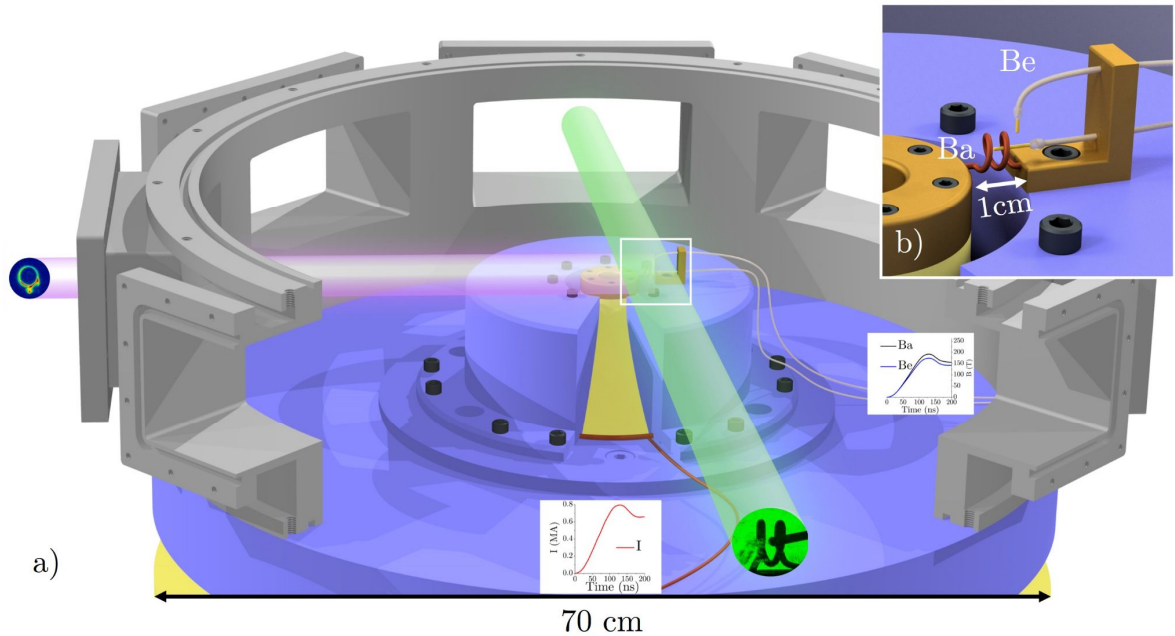

Figure 1. a) Overall view of the experimental setup showing the XUV line of sight (purple) and the laser shadowgraphy line of sight (green). The Rogowski coil (brown) that measures the discharge current is located at the base of the anode cap (blue), near the cathode (yellow). The two Bdot probes are to the right the anode cap. b) A enlarged view of the load area showing the coil and location of the edge (Be) and axial (Ba) Bdot probes.

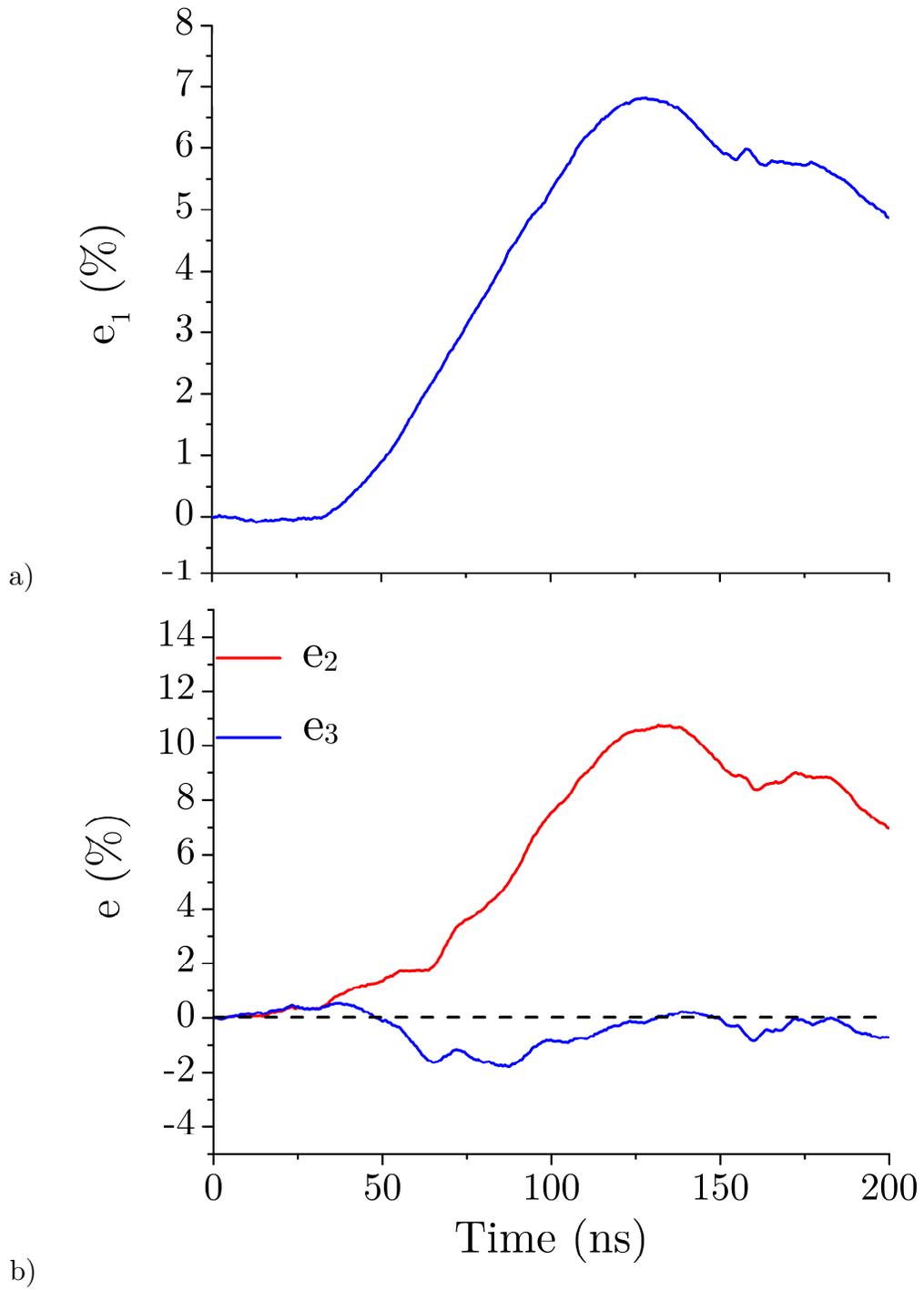

Figure 2. a) The error $e_1$ for a ratio $r_1$ of 7.4. b) the error $e_2$ and $e_3$.

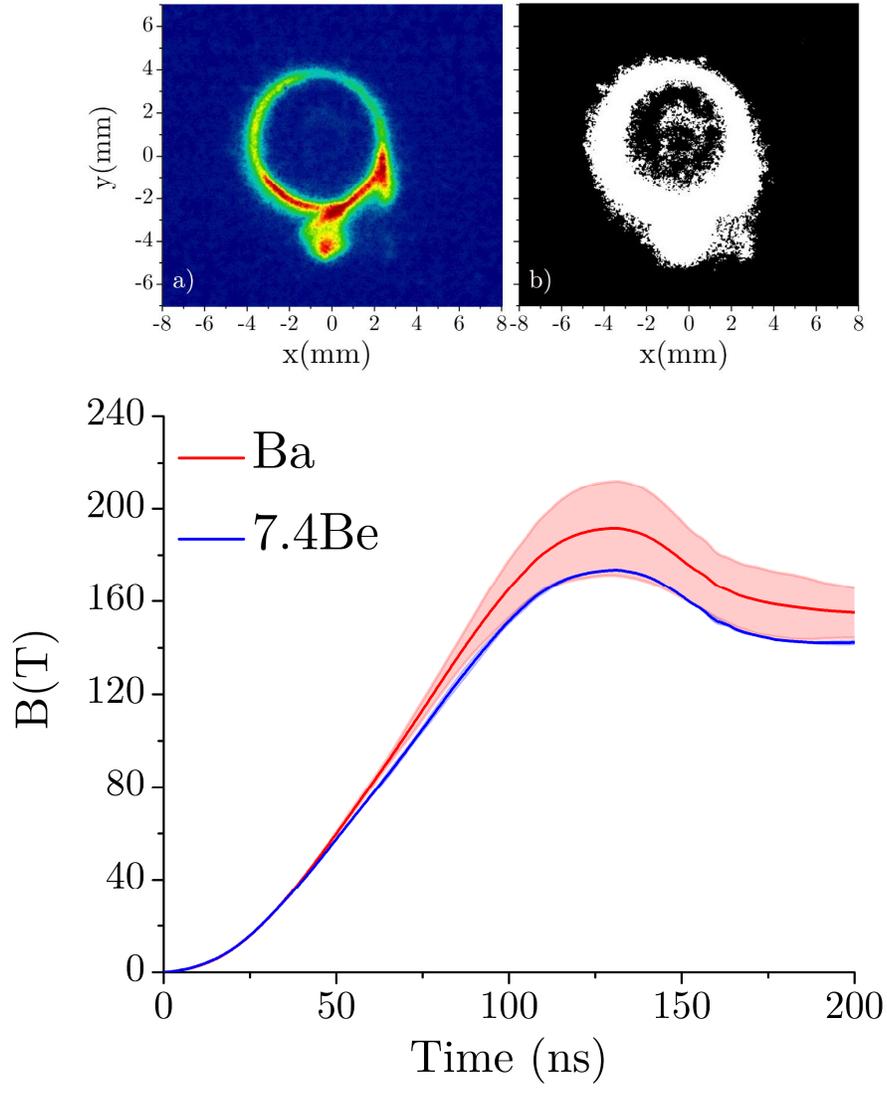

**Figure 3.** a) End-on view of the self-emission in the XUV spectrum for the initial calibration case, using an arbitrary color scale, 170 ns into the current discharge. b) The same view only for light intensity renormalized to the probe self-emission. c) Ba and 7.4Be as a function of time in a detectable plasma environment. The shaded areas represent the errors computed using Eq. (3) for Ba and Be respectively.

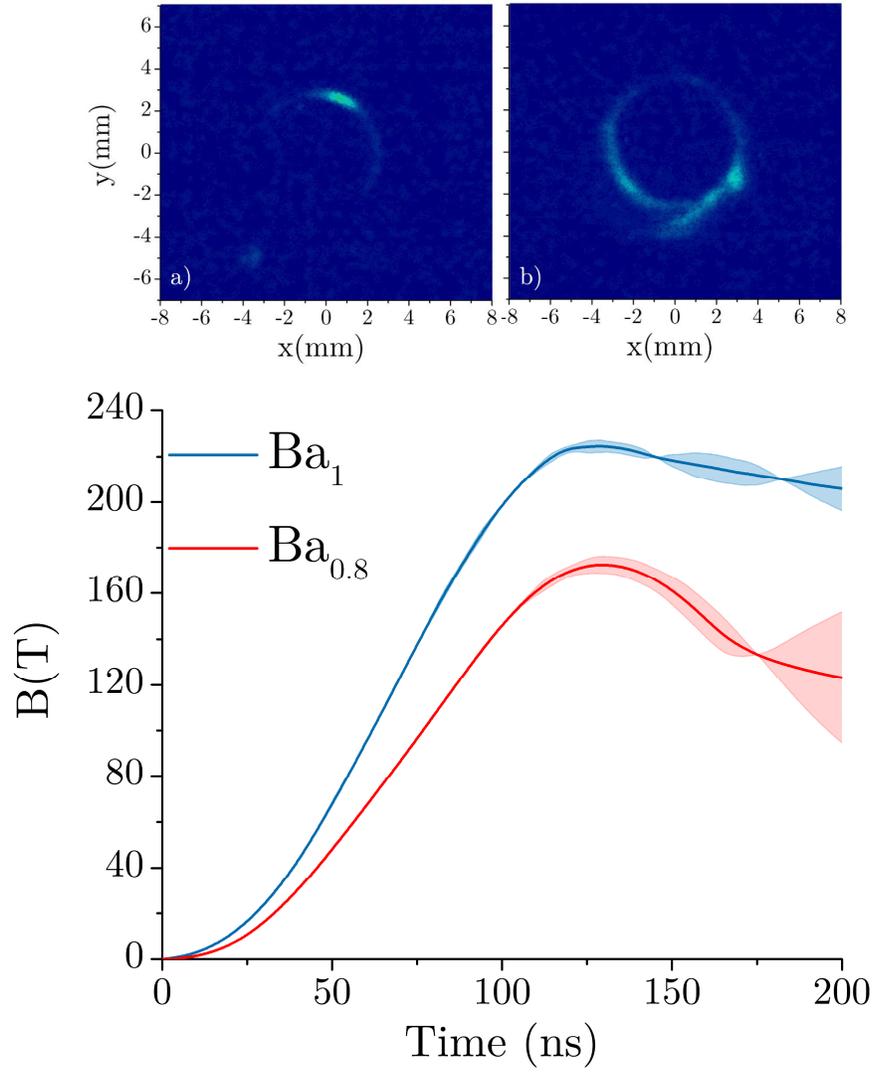

**Figure 4.** End-on view of the self-emission in the XUV spectrum at current peak, shown in false colors, using the same scale as in Figure 3, for a) 810 kA ($Ba_{0.8}$) and b) 1.04 MA ($Ba_1$). c) Ba as a function of time for both currents. The shaded areas represent the errors computed using Eq. (3).

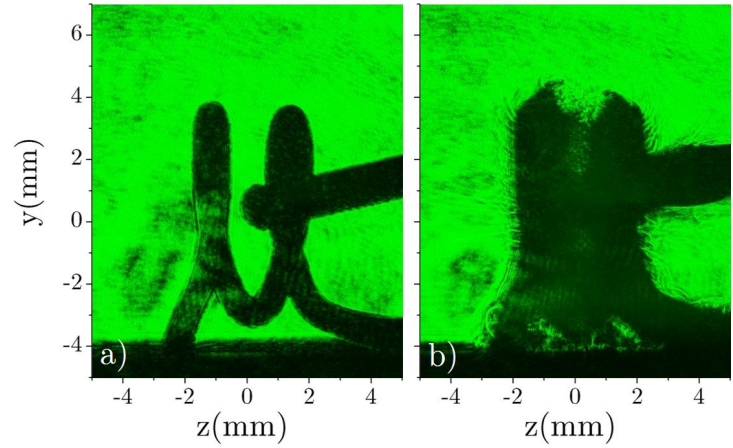

**Figure 5.** Laser backlighter showing the initial coil setup with Bdot probe a) before the shot and b) 246 ns into the current discharge ($I_{peak}$=700 kA $t_{rise}$= 130 ns), a time at which the plasma generated between two consecutive turns is dense enough to cut-off green laser light.

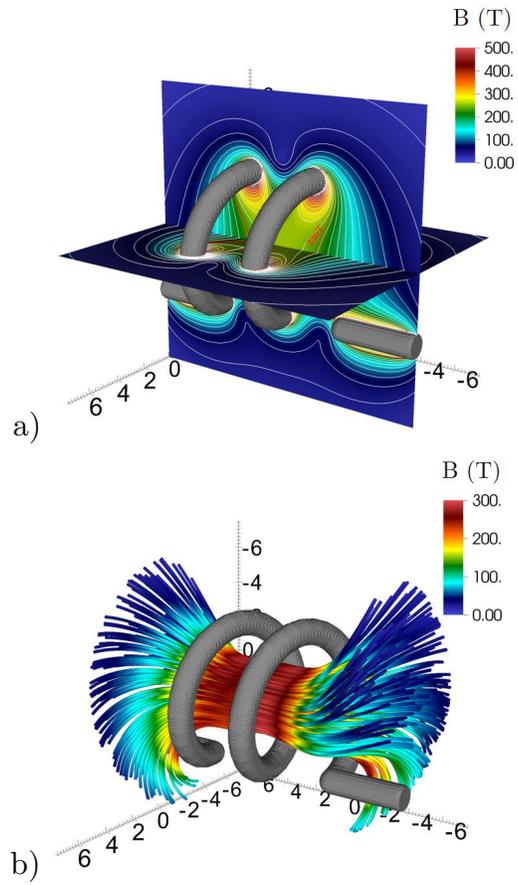

**Figure 6.** Computation of a) field strength (with white iso-contours every 20 T) and b) field lines for the coil used experimentally using Biot and Savart's law.

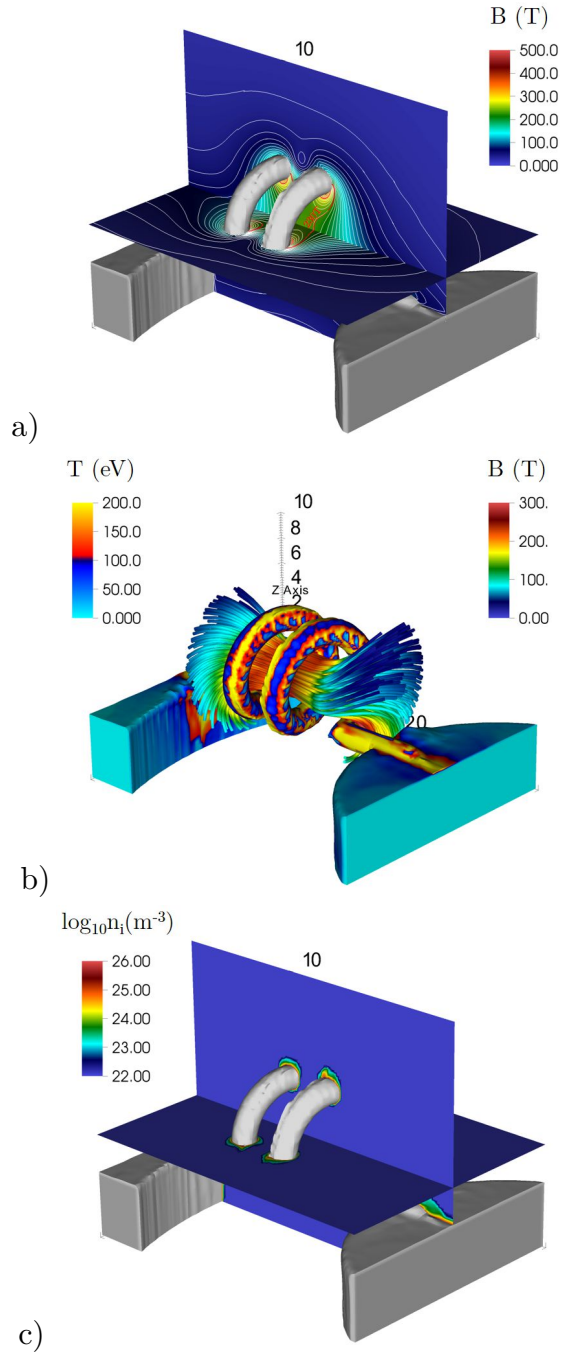

**Figure 7.** Computation of a) field strength (with white iso-contours every 20 T) and b) field lines for the coil used experimentally using PERSEUS. c) Most of the volume is devoid of plasma with density larger than $10^{22}$m$^{-3}$. Sizes are in millimeters. The density cut-off for the coil, anode and cathode is $10^{25}$m$^{-3}$.